\documentclass[12pt,preprint,epsfig]{aastex}
\usepackage[T1]{fontenc}
\usepackage[latin1]{inputenc}
\usepackage{amsmath}
\usepackage{graphics}

\makeatletter

\providecommand{\LyX}{L\kern-.1667em\lower.25em\hbox{Y}\kern-.125emX\@}
\let\SF@@footnote\footnote
\def\footnote{\ifx\protect\@typeset@protect
    \expandafter\SF@@footnote
  \else
    \expandafter\SF@gobble@opt
  \fi
}
\expandafter\def\csname SF@gobble@opt \endcsname{\@ifnextchar[
  \SF@gobble@twobracket
  \@gobble
}
\edef\SF@gobble@opt{\noexpand\protect
  \expandafter\noexpand\csname SF@gobble@opt \endcsname}
\def\SF@gobble@twobracket[#1]#2{}

\shorttitle{Relativistic observables for PSR~J1141--6545}
\shortauthors{Bailes et al.}
\makeatother

\begin{document}

\title{
Self-consistency of relativistic observables with general
relativity in the white dwarf-neutron star
binary pulsar PSR~J1141--6545.
}

\renewcommand{\familydefault}{bch}

\topmargin=0in        
\oddsidemargin=0in    
\evensidemargin=0in   
\textwidth=6.5in      
\textheight=8.5in     

\headheight=4mm
\headsep=4mm

\newcommand{\bsf}{\sffamily\bfseries}

\newcommand{\msun}{\mbox{M$_{\odot}$}}
\newcommand{\rsun}{\mbox{R$_{\odot}$}}
\newcommand{\aopx}{\mbox{$\Delta_{\pi \rm M}$}}
\newcommand{\shap}{\mbox{$\Delta_{\rm S}$}}
\newcommand{\psr}{\mbox{PSR J1141$-$6545\,}}
\newcommand{\mcrit}{\mbox{$M_{\rm crit}\,$}}
\newcommand{\mpsr}{\mbox{$M_{\rm p}\,$}}
\newcommand{\mwd}{\mbox{$M_{\rm c}\,$}}

\newcommand\wvect[1]{{\bsf #1}}
\newcommand\bvect[1]{{\bsf #1}$_{\bm 0}$}

\newcommand\etal{et al.}

\def\degr{\ifmmode^{\circ}\else$^{\circ}$\fi}
\def\agep{\ifmmode\tau_{\rm c}\else$\tau_{\rm c}$\fi}
\def\agec{\ifmmode\tau_{\rm w}\else$\tau_{\rm w}$\fi}

\affil{Centre for Astrophysics and Supercomputing, Swinburne
University of Technology, P.O.~Box 218 Hawthorn, VIC 3122, Australia
mbailes@swin.edu.au}

\author{ M. Bailes,
	S. M. Ord,
	H. S. Knight,
	A. W. Hotan}
 
\begin{abstract}

Here we report timing measurements of the relativistic binary pulsar
\psr\, that constrain
the component masses and demonstrate that the orbital
period derivative $\dot P_{\rm b}= (-4\pm 1)\times 10^{-13}$ 
is consistent with gravitational wave emission as
described by the general theory of relativity.
The mass of the neutron star and its companion
are 1.30$\pm$0.02 \msun\, and 0.986$\pm$0.02 \msun\, respectively,
suggesting a white dwarf companion, and extending the range of
systems for which general relativity provides a correct
description. On evolutionary grounds,
the progenitor mass of \psr \,should be near the
minimum for neutron star production. Its mass
is two standard deviations below the mean of
the other neutron stars, suggesting a relationship
between progenitor and remnant masses.
\end{abstract}


\section{Introduction}

Our knowledge of 
neutron star masses \cite{tc99} is derived largely
from three relativistic binary
pulsars with eccentric orbits and probable neutron star
companions \cite{wt03,sttw02,dk96}. The masses
of all known neutron stars exhibit a remarkably
narrow scatter (0.04\msun) about their mean of 1.35 \msun.
If one restricts attention to the neutron star pairs, the
mean is 1.37 \msun.
These binary pulsars have also been pivotal in tests of
the general theory of relativity (Taylor, Fowler and
McCulloch 1979)\nocite{tfm79}.

Gravitational waves have never been observed directly. Our
best quantitative evidence for their existence comes from observations
of the binary pulsar PSR B1913+16 \nocite{dt91}
(Damour and Taylor 1991). 
This pulsar has a negative orbital period
derivative that is consistent with the emission of
gravity waves at the 0.4\% level.
Unfortunately, the binary 
PSR B1534+12 is so close to the Earth
that its proper motion induces a time-varying
Doppler term into the timing which prevents separation
of the relativistic contribution to the orbital
period derivative from the kinematic one \cite{bb96,sttw02}. 
Until an independent distance
can be established, this pulsar's orbital period
derivative cannot verify the predictions of the 
theory to better than $\sim 20$\%. 
Nevertheless, measurement of PSR B1534+12's post-Keplerian
parameters have provided other tests of general
relativity \cite{sttw02}, and binary pulsars such
as PSR J0437--4715 have proved that the theory is
consistently in agreement with the most precise
observations \cite{vbb+01}.

Discovered in the Parkes multibeam survey \nocite{klm+00a}
\psr\, is a 4.7$^{\rm h}$ binary pulsar with an eccentricity
of 0.17 that consists of a 394 ms relatively
youthful pulsar of characteristic age $\sim $1.4 Myr and an
unidentified companion star. Early observations \cite{klm+00a} 
of the periastron advance suggested a pulsar mass $\mpsr<
1.35$ \msun, and a companion $M_{\rm c}>0.97$ \msun. The
observed tight distribution of neutron star masses near
1.35$\pm$0.04 \msun\, where the quoted error reflects a Gaussian
fit to the observed distribution,
suggested that this system probably
consisted of a $\mpsr\sim$ 1.3 \msun\, neutron star and 
$\mwd \sim$ 1 \msun\, white dwarf companion. \psr\, was
therefore unique among the relativistic binary pulsars. Its 
component masses argued for a white dwarf companion
yet its orbit was highly eccentric making it the first
of a new class of object. To date no white dwarf has
been optically identified.

\psr\, is an exciting target for timing.
Firstly, scintillation observations have demonstrated a
clear modulation of the timescale of the interstellar
``twinkling'' allowing derivation of the system's 
angle of inclination and runaway velocity (Ord, Bailes
and van Straten 2002a)
\nocite{obv02}.
Two degenerate fits to the changing scintillation timescale
were possible. The observed distribution of neutron
star masses favoured the solution with
an inclination angle
($i=76\pm 2.5\degr$) and center of mass velocity
in the plane of the sky of $V\sim 115$ km s$^{-1}$. These
results provide an independent estimate of the
system orientation to be compared with the timing.
Secondly,
neutral hydrogen observations (Ord, Bailes and van Straten 
2002b) \nocite{obv02b} have demonstrated that this
pulsar is in a favourable location in the Galaxy
where the observed orbital period derivative will, as in
the case of PSR B1913+16, 
be dominated by the general relativistic contribution.
Finally, contrary to PSR B1913+16, which is a very
symmetric system with two neutron stars
of nearly equal mass, \psr\, is a very dissymmetric
system as it comprises a strongly self-gravitating
neutron star and a (relatively) weakly self-gravitating
white dwarf. Timing this system therefore tests
new aspects of relativistic theories of gravity \cite{de92,twdw92,
arz03}. 

\begin{figure}[here]
\label{fig:geometry}
\plotone{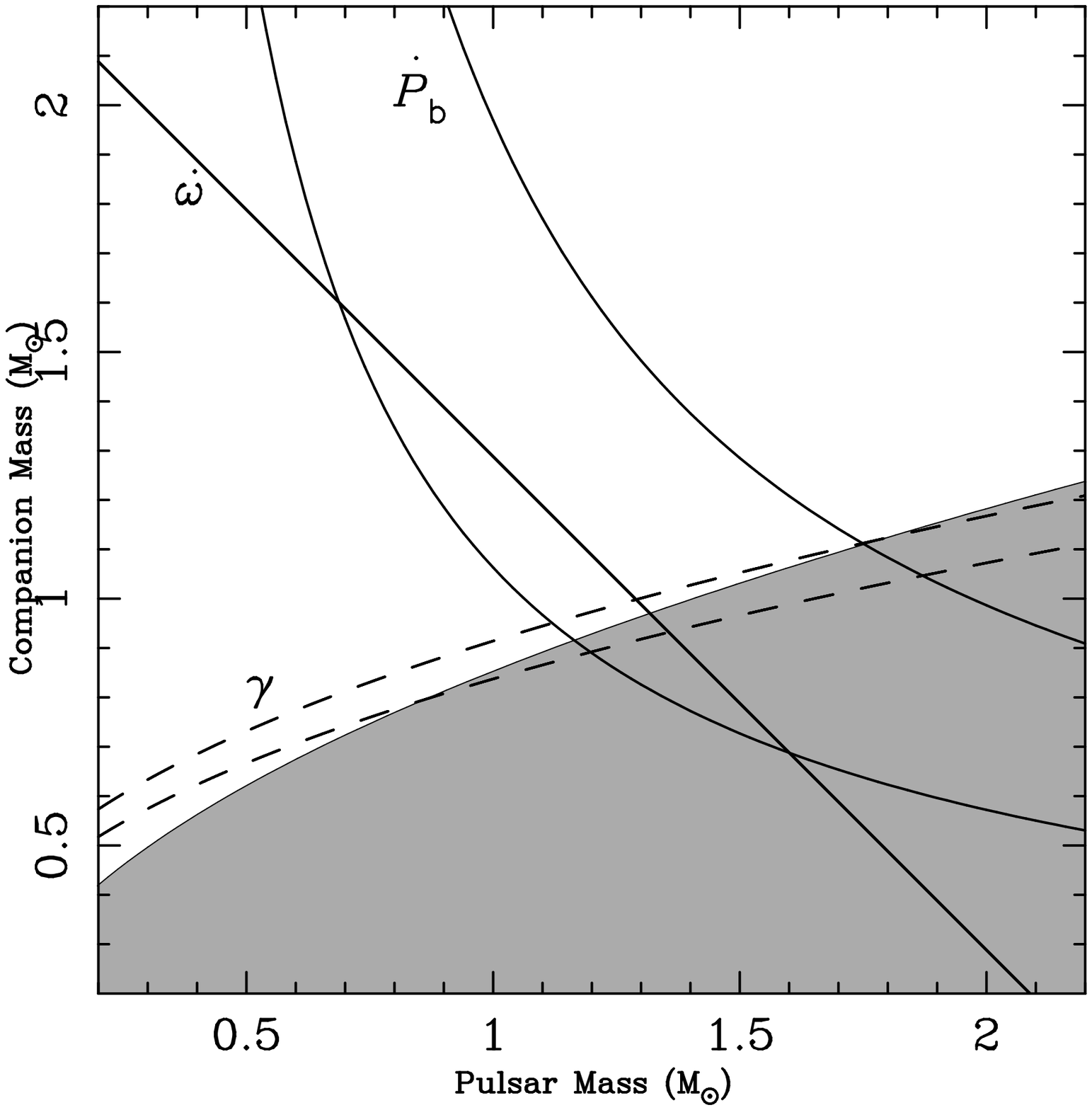}
\caption{
Constraints on the pulsar and companion masses. The shaded area
is excluded from a consideration of Kepler's laws. The sum
of the masses $M=2.2883\pm 0.0006 \msun$ is very accurately 
defined from the relativistic
advance of periastron, $\dot{\omega}$, and is shown by the straight solid line
with a slope of --1.
The relativistic term $\gamma$ constrains the system to lie
between the dashed lines, and the measured orbital period
derivative $\dot P_{\rm b}$ suggests that the system masses
lie between the pair of curved solid lines.
This implies an inclination angle $i>75\degr$, consistent
with constraints from scintillation studies $i=76\pm2.5\degr$.
The derived component masses suggest 
that the system consists of a 0.986$\pm$0.020 \msun white dwarf and a 
1.30$\pm$0.02 \msun pulsar. If the system radiates energy
at the rate predicted by general relativity,
the measured orbital
period derivative $\dot P_{\rm b} = (-4\pm1) \times 10^{-13}$
is consistent with the predicted value of ($-3.8\times10^{-13}$)
derived from the component masses. }
\end{figure}

\section{Observations and Data Analysis}

Observations of \psr\ were conducted from January 2001 until February
2003 at the Parkes 64\,m radio telescope as part of a scintillation
study.  The data from two orthogonal linear polarisations were
detected and summed using a multi-channel filterbank consisting of
2$\times$512$\times$0.5 MHz filters.  These data were 1-bit sampled
every 125 $\mu$s and written to magnetic tape for offline processing
on the Swinburne University of Technology supercomputer.  Observations
were usually greater than one orbit in duration, and there were
ten sessions, nine of which took place in the 15 months prior to
Feb 2003.  The central
observing frequency was 1390 MHz and total intensity pulse profiles
were formed by folding the raw data at the topocentric period of the
pulsar for integration times of 60\,s. 

A fit to the arrival times was made for the standard pulsar spin,
astrometric, and Keplerian terms as well as three post-Keplerian
parameters using a relativistic model \nocite{dd86}
(Damour and Deruelle 1986).  After incorporation
of archival timing data from the Australia Telescope National Facility
archives from July-August 1999 we found that our measurements of the
periastron advance $\dot \omega$, transverse Doppler and gravitational
redshift, $\gamma$ term, and orbital period derivative $\dot P_{\rm
b}$ became much more significant. These values are displayed in Table
1.  The archival data consisted mainly of data taken with a
2$\times$96$\times$3 MHz filterbank system. We fit for a single
arbitrary offset between the two datasets to allow for the change in
instrumentation.

\section{Interpretation}

Newton's laws provide a relation between the system's
masses and inclination angle $i$, and the common observables
orbital period $P_{\rm b}$ and projected semi-major axis $x = (a \sin i)/c $.
This is often referred to as the pulsar mass function $f_{\rm P}$,
\begin{equation}
f_{\rm P}=
{{(\mwd \sin i)^3}\over{M^2}} = 
\Big({{2\pi}\over{P_{\rm b}}}\Big)^2
{{(xc)^3}\over{G}}
\end{equation}
here $M$ is the sum of the masses, $G$ is Newton's gravitational
constant and $c$ is the speed of light. \mpsr and \mwd refer
to the mass of the pulsar and companion star respectively.

General Relativity allows us to use the post-Keplerian terms
to derive component masses. The advance of periastron
gives us the sum of the masses.
\begin{equation}
\dot \omega = 3 \Big({{2 \pi}\over{P_{\rm b}}}\Big)^{5/3}
\Big({{G M}\over{c^3}}\Big)^{2/3}
(1-e^2)^{-1}
\end{equation}
Here $e$ is the orbital eccentricity.
The combined transverse Doppler and gravitational
redshift term gives a different relation between
the masses. 
\begin{equation}
\gamma = e \Big({{P_{\rm b}}\over{2\pi}}\Big)^{1/3}
G^{2/3}c^{-2}\mwd(\mpsr+2\mwd)M^{-4/3}.
\end{equation}
General relativity predicts that gravitational
wave emission from binary systems removes
energy that results in an orbital decay.
The system is over-determined if one can also measure
the orbital period derivative $\dot P_{\rm b}$. 
\begin{displaymath}
\dot P_{\rm b} = 
- {{192 \pi G^{5/3}}\over{5 c ^5}}
\Big({ {2\pi}\over{P_{\rm b}}}\Big)^{5/3}(1-e^2)^{-7/2}
\end{displaymath}
\begin{equation}
\times  \Big(1+{{73}\over {24}}e^2
+{{37}\over{96}} e^4\Big)\mpsr \mwd M^{-1/3}
\end{equation}
Thus, binary pulsars such as \psr can be used to
test the validity of the general theory of relativity.
The allowed range of masses for the system are shown in Figure 1.
Other post-Keplerian parameters such as the range and shape
of the Shapiro delay were not significant in the fit.

The companion is almost certainly a white dwarf, unless
neutron star masses extend over a very large range.
The mean and standard deviation of the 7 neutron stars
shown in Figure 2 would place the companion of \psr
seven standard deviations from the mean of the others which
appears unlikely.

Young pulsars often exhibit
significant timing noise which make timing proper motions
unreliable on the timescale of a few years. A formal
fit to the limited data yielded proper motions
in both right acension and declination of --12$\pm$3 milli-arc
seconds per year. Whether
the apparently 4-sigma proper motion is real or not
will become apparent with continued timing. For now
we are treating the result with some caution. Timing
noise has a much longer period (months to years) than
the orbital period of our pulsar (4.7h), and so
although the timing proper motion is potentially non-physical,
the derived orbital parameters are not subject to contamination
by timing noise.

The self-consistency of the system with general relativity
is very pleasing. The orbital period derivative
is now a 4-sigma result. The relative
error in this term scales as the length of the 
observing span $t^{-5/2}$, so by the
end of the decade it should be near 1\%.
\psr is therefore likely to be a very important
new testing ground for general relativity. The dissymetry
of the system tests different aspects of the
theory since one of the stars is not strongly
self-gravitating.
Its favourable location in the
Galaxy means that the measured orbital period
derivative is close to the intrinsic value (Ord, Bailes
and van Straten 2002a). Unfortunately, the pulsar
is young, and unlikely to be as intrinsically stable as PSR B1913+16,
which has a much weaker magnetic field strength.
Whether this is important or not for post-Keplerian orbital parameters
remains to be seen. In any case, it will be difficult to
surpass PSR B1913+16's relative errors on terms such
as the relativistic contribution to the orbital period
derivative.

\section{Discussion}

Large-scale simulations of binary pulsar progenitors
predict a large population of eccentric pulsar-white dwarf 
binaries \cite{ts00a},
albeit with a short observable lifetime. 
\psr is thought to arise from the
evolution of a binary in which both stars are initially
less than the critical mass \mcrit required to produce a supernova
\cite{ts00a}.
The initially more-massive star transfers mass to its companion
before becoming a white dwarf. If sufficient matter can
be accreted by the initially lower-mass star it will exceed \mcrit
and produce a pulsar. Should the system remain
bound, we will be left with a
young neutron star orbiting a white dwarf companion in
an eccentric orbit. \psr's mass \mpsr is therefore very important,
because we know that the progenitor must have been less
than 2 \mcrit -- \mwd. This is the first time we have an
upper limit to the progenitor mass of a neutron star, albeit
in terms of \mcrit. 
All previous neutron star mass measurements are from
systems where both components were large
enough to form neutron stars, which makes the progenitor masses
far more difficult to constrain as our theoretical 
understanding of the mass of neutron star progenitors is
limited.

Radio pulsar masses cannot be easily determined.
Neutron stars can theoretically exist over a range of 
masses 
(Timmes, Woosley and Weaver 1996)
\nocite{tww96} that is much broader than that
inferred from observations of radio pulsars \nocite{wt03,
sttw02,dk96}(
Weisberg and Taylor 2003, Stairs \etal 2002, Deich and
Kulkarni 1996).
Until these measurements, the mean and standard
deviation of the six accurately known neutron stars
was just 1.37$\pm$0.04 \msun. 
\psr's mass of 1.30$\pm$0.02 is the smallest yet 
determined with any accuracy. It's mass is
two standard deviations below the mean 
of the others as shown in Figure 2. 

\begin{figure}[here]
\label{fig:junk}
\epsscale{1.0}
\plotone{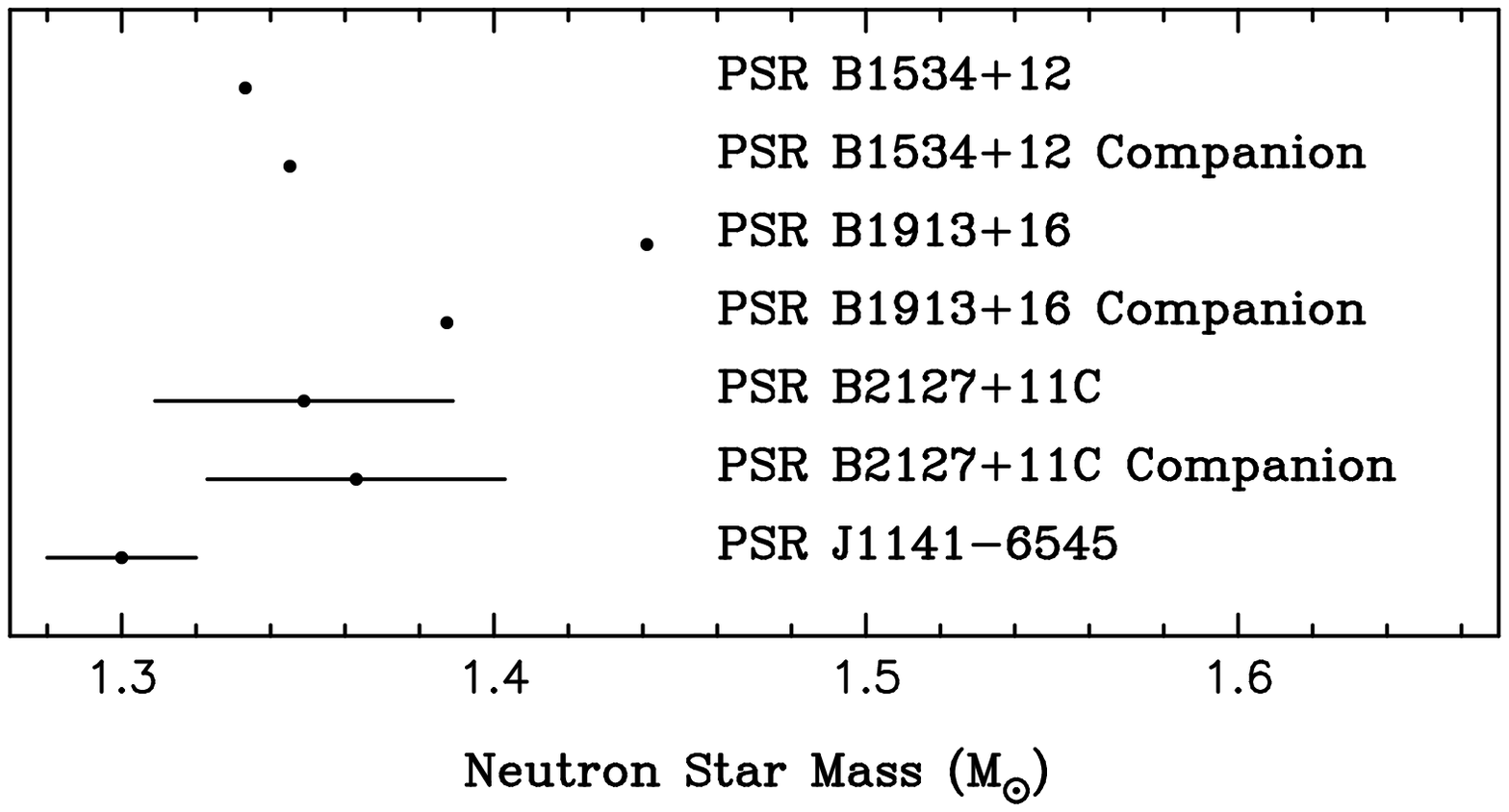}

\caption{
Neutron star masses derived from relativistic binary pulsars.
Error bars are shown for all data, but in the case of
PSR B1913+16 and PSR B1534+12 they are too small to be visible.
Masses are taken from Deich and Kulkarni (1996),
Stairs {\it et al.} 2002 and Weisberg and Taylor (2003).
\psr\, is the first neutron star not in a neutron star-neutron star
binary
to have its mass so precisely determined.
On evolutionary grounds we know its progenitor was near 
the lower end of
the mass distribution for pulsar progenitors.
The low remnant mass suggests that the neutron star
mass is related to that of the progenitor.
}
\end{figure}

\noindent
In the absence of any asymmetry in the explosion,
component masses and orbital parameters define the
progenitor system mass and runaway velocity \cite{rs86}.
If the explosion were symmetric, the pre-supernova star in 
the \psr\, system would
have been only $M_{\rm pre}=M(1+e)-\mwd=1.7$ \msun. It seems unlikely that
the presupernova star could be such a small fraction
of its original mass $> \mcrit \sim 8-11$ \msun.
Thus \psr's orbital configuration is therefore a
strong argument in favour of asymmetric supernova kicks
\cite{dc87,bai89}.
Such kicks can misalign the spin and orbital angular
momentum vectors of the pulsar, allowing detectable
precession due to spin orbit coupling\cite{dr74,bo75}.
This has been observed in the binary pulsar PSR 1913+16 \cite{kra98}.
In \psr\, the timescale for complete precession is
only $\sim$ 200 yr. Its high orbital inclination
could lead to significant changes in our line of sight to
the pulsar's spin axis on the timescale of a decade or so.
This might explain the non-detection of the pulsar
in previous surveys of the region and we would
expect rapid evolution of the pulse profile as
suggested by \cite{klm+00a}.

\psr\, is a very significant object. It
has already validated scintillation speed methodologies,
provided the first estimate of a neutron star mass
where we have firm constraints on the progenitor,
proved that mass transfer can transform a star
that would otherwise become a white dwarf into a
neutron star and extended the range of systems
for which general relativity provides an accurate
description. Its orbital parameters 
suggest that the explosion was probably asymmetric.
As the errors in orbital period
derivative decrease, this system will test
new aspects of theories of relativistic gravity that
were impossible with the other relativistic binaries
due to its unique composition 
\cite{twdw92,dt92}. Together, PSR B1913+16, PSR B1534+12
and PSR J1141--6545 will strongly constrain relativistic
theories of gravity.

\bibliographystyle{apj00}

\vspace{5mm}
\noindent
\bsf{Acknowledgements} \\

\noindent
{\small The Parkes Observatory is part of the Australia Telescope
which is funded by the Commonwealth of Australia for operation as a
National Facility managed by CSIRO. We thank Dick Manchester for
his prompt supply of archival timing data on this pulsar
and the multibeam collaboration for use of their equipment,
T. Damour for comments on an earlier version of the manuscript,
and the referee D. Nice for helpful suggestions.
M.~Bailes is an ARC Senior Research Fellow. H. Knight \&
A. Hotan are recipients of CSIRO studentships.
\\
}

\begin{table}
\begin{center}
\begin{tabular}{lr}
\hline
\multicolumn{2}{c}{\bf\textsf{Table 1 Observed and derived parameters}} \\
\hline \\
Right ascension, $\alpha$ (J2000)       \dotfill &
			11$^{\mathrm h}$41$^{\mathrm m}$07\fs022(6) \\
Declination, $\delta$ (J2000)           \dotfill & 
			-65\degr45\arcmin19\farcs089(9) \\
Pulse period, $P$ (ms)                    \dotfill & 393.8978340370(2) \\
Reference epoch (MJD)                     \dotfill & 51369.8525  \\
Dispersion Measure cm$^3$ pc$^{-1}$ \dotfill & 116.048(2)\\
Period derivative, $\dot{P}$ (10$^{-15}$) \dotfill & 4.294593(3) \\
Orbital period, $P_{\rm b}$ (days)        \dotfill & 0.1976509587(3) \\
Projected semi-major axis $x$ (s)        \dotfill & 1.85894(1) \\
Orbital eccentricity, $e$        	  \dotfill & 0.171876(2) \\
Epoch of periastron, $T_{0}$ (MJD)        \dotfill & 51369.854553(1) \\
Longitude of periastron, $\omega$ (\degr) \dotfill & 42.457(2) \\
Number of arrival times \dotfill & 7180\\
RMS Residual ($\mu$s) \dotfill & 114\\
Damour-Deruelle PK parameters\\
$\gamma$ (s) \dotfill & 0.00072(3)\\
$\dot \omega$ (\degr yr$^{-1}$)           \dotfill & 5.3084(9) \\
$\dot P_{\rm b} (10^{-12})$      	  \dotfill & --0.43(10) \\

Damour-Deruelle GR fit\\

Companion mass, $\mwd$ (M$_{\odot})$       \dotfill & 0.986(20) \\
Pulsar mass, $\mpsr$ (M$_{\odot})$       \dotfill & 1.30(2) \\
Sum of masses, $M$ (M$_{\odot})$ \dotfill & 2.2883(5)\\
Derived Parameters\\
Orbital inclination, $i$ (\degr)          \dotfill & $i>75$ \\

\end{tabular} \\

\vspace{5mm}
\parbox{88mm}{\sffamily 
Best-fit physical parameters and their formal 1$\sigma$
errors were derived from arrival time data by minimizing an objective
function, as implemented in TEMPO
(http://pulsar.princeton.edu/tempo). The timing model used is
TEMPO's DD relativistic binary model \cite{dd86} which
allows us to fit for $\dot \omega$, $\dot P_{\rm b}$ and
$\gamma$ separately.
To derive the best estimate of the companion and pulsar
masses, a fit to only $M$ and \mwd was performed using TEMPO's
DDGR model. This model assumes general relativity is correct.
Parenthesized numbers represent
the 1-$\sigma$ uncertainty in the last digits quoted, and epochs are specified using
the Modified Julian Day (MJD). 
}
\end{center}
\end{table}

\end{document}